\documentclass[final,5p,times,twocolumn]{elsarticle}
\usepackage{graphicx}
\usepackage{amssymb}
\journal{PHYSICA E}
\begin{document}

\begin{frontmatter}

\title{Indications for a line of continuous phase transitions at finite
temperatures \linebreak connected with the apparent metal-insulator 
transition in 2d disordered systems}

\author{Arnulf M\"obius}

\address{Leibniz Institute for Solid State and Materials Research
IFW Dresden, \linebreak POB 270116, D-01171 Dresden, Germany}

\ead{a.moebius@ifw-dresden.de}

\begin{abstract}
In a recent experiment, Lai {\it et al.}\ [Phys.\ Rev.\ B {\bf 75} 
(2007) 033314] studied the apparent metal-insulator transition (MIT) 
of a Si quantum well structure. Tuning the charge carrier concentration 
$n$, they measured the conductivity $\sigma(T,n)$ for a very dense set
of $n$ values. They observed linear $T$ dependences of $\sigma$ around 
the Fermi temperature and found that the corresponding $T \rightarrow 0$
 extrapolation $\sigma_0(n)$ exhibits a sharp bend just at the MIT. 
Reconsidering the data by Lai {\it et al.}, it is shown here that this 
sharp bend is related to a peculiarity of $\sigma(T = {\rm const.},n)$, 
which is clearly detectable in the whole $T$ range up to $4\ {\rm K}$, 
the highest measuring temperature in that work. It may indicate a sharp 
continuous phase transition between the regions of apparent metallic and
activated conduction to be present at finite temperature. This 
interpretation is confirmed by a scaling analysis without fit, which
illuminates similarities to previous experiments and provides 
understanding of the shape of the peculiarity and of sharp peaks found 
in $\mbox{d}\;\!\mbox{log}_{10}\;\!\sigma / \mbox{d}\;\!n\,(n)$.
Simultaneously, the scaling analysis uncovers a strange feature of the 
apparent metallic state.
\end{abstract}


\end{frontmatter}

\section{Introduction}

Is electronic transport in 2-dimensional (2d) disordered systems 
exclusively nonmetallic, so that the resistivity $\rho(T)$ always 
diverges as the temperature $T$ tends to 0, or can it also have metallic 
character? This fundamental problem has been controversially debated
for three decades: The existence of a corresponding metal-insulator
transition (MIT) was denied by the localisation theory of Abrahams {\it
et al.}\ \cite{Abra.etal.79}, which, however, neglects electron-electron
interaction. Thus, it came as a surprise when Kravchenko {\it et al.},
who had varied the charge carrier concentration $n$ in high mobility
MOSFET samples, first reported a strong decrease of $\rho$ down to
$20\ {\rm mK}$ \cite{Krav.etal.94}. They considered the conduction in 
the respective $(T,n)$ region as metallic, an interpretation which was
questioned in particular by Altshuler and Maslov \cite{Alts.Masl.99}.
The nature of the apparent metallic state has remained puzzling up to 
now, for recent reviews see Refs.\ \cite{Krav.Sara.04,Puda.04}.

Recently, Lai and coworkers studied the transport properties close
to the apparent MIT of an n-type Si quantum well confined in a
Si$_{0.75}$Ge$_{0.25}$/Si/Si$_{0.75}$Ge$_{0.25}$ heterostructure 
\cite{Lai.etal.07}. The authors observed the conductivity $\sigma$ to 
have nearly linear $T$ dependences around the Fermi temperature 
$T_{\rm F}$ varying between 2 and $2.5\ {\rm K}$. They showed that 
$\sigma_0(n)$, the $T=0$ conductivity obtained by linear extrapolation 
from this $T$ region, exhibits two regimes of different slope. 
Particularly important, there is a sharp bend at the transition between 
the two regimes. It coincides with the $n$ value, $n_{\rm c}$, where 
$\mbox{d}\:\!\sigma / \mbox{d}\:\!T$ changes its sign as $T \rightarrow 0$. 
The authors interpret this finding as an indication of the existence of 
two different phases. Thus, the apparent MIT should be related to a real
phase transition at $T = 0$. This transition is commonly expected to be 
smoothed out at finite $T$. 

However, such a smoothing is not a must for a quantum phase transition
as is obvious from two examples: The Ising ferromagnet LiHoF$_4$ in 
transverse magnetic field undergoes a quantum phase transition at 
$T = 0$, where the quantum critical point is the beginning of a line of 
continuous phase transitions at finite $T$ \cite{Bitk.etal.96}. The 
phase diagram of the quasi-two-dimensional organic conductor 
$\kappa$-(BEDT-TTF)$_2$X exhibits a line of first-order transitions at 
finite $T$ separating an antiferromagnetic-insulator phase from an
unconventional-superconductor phase at low $T$ \cite{Kaga.etal.05}.

From this perspective, the additional observation by Lai {\it et al.}, 
that the slope of $\sigma(T)$ for $T \sim T_{\rm F}$ is almost constant 
close to the MIT, $n = n_{\rm c}$, provokes two interesting questions: 
(i) Provided the extrapolated $\sigma_0(n)$ has a knee indeed, and the 
slope used in extrapolation is roughly constant, should not 
$\sigma(T = {\rm const.},n)$ exhibit a knee at $n_{\rm c}$ also for 
finite $T$? (ii) If yes, is the existence of this peculiarity restricted
to the region of linear $\sigma(T)$ around $T_{\rm F}$, or is it a more 
general phenomenon?

Starting from these questions, the present work points to several 
indications for the MIT in 2d systems being connected with 
a line of continuous phase transitions at finite $T$. It substantially 
extends the detailed arguments given in Ref.\ \cite{Moe.08}: More data 
are taken into consideration in the scaling region, where
$\sigma(T,n) = \sigma((n-n_{\rm c})/T^{1/\beta})$, consequences of this 
scaling for the physical nature of the apparent metallic state are 
discussed, and possible corresponding similarities with the transport in
an AlAs quantum well \cite{Papa.Shay} are illuminated.

\section{Phenomenological reconsideration of the experiment by Lai and
coworkers}

To find out to which extent and in which form the sharp bend of
$\sigma_0(n)$ persists when $T$ is finite, I digitised the 
high-resolution preprint version of Fig.\ 1(b) of Ref.\ 
\cite{Lai.etal.07}. Fig.\ 1 depicts the obtained 
$\sigma(T = {\rm const.},n)$ data for five $T$ values. First of all, 
this graph shows that all $\sigma(T = {\rm const.},n)$ curves share a 
peculiar feature, namely an ``indentation'' at 
$n \approx 0.32 \cdot 10^{11}\ {\rm cm}^{-2}$. It occurs close to the 
concentration value where $\mbox{d}\:\!\sigma / \mbox{d}\:\! T$ changes 
its sign at the lowest $T$ considered in Ref.\ \cite{Lai.etal.07}. This 
value, $n_{\rm c} = 0.322 \cdot 10^{11}\ {\rm cm}^{-2}$, is the critical 
concentration of the apparent MIT. It is marked by arrows in Figs.\ 1, 
2, and 4.

\begin{figure}
\begin{center}
\includegraphics[width=0.70\linewidth]{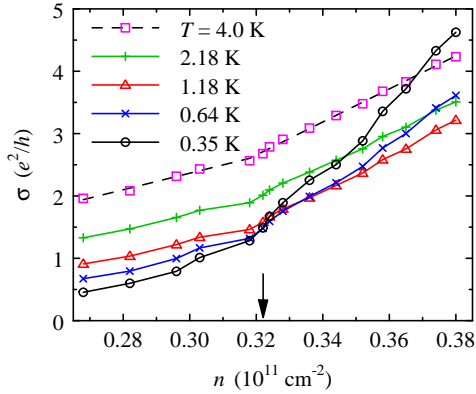}
\end{center}
\caption{Charge carrier concentration dependence of the conductivity for
the Si quantum well structure studied in Ref.\ \cite{Lai.etal.07}. Data
were obtained from Fig.\ 1(b) of that work. The arrow marks the critical
concentration $n_{\rm c}$ of the apparent MIT, see text. The dashed line
is a fit by the piecewise linear function introduced in the text. The 
full lines serve as guide to the eyes.}
\end{figure}

To answer the question whether or not the sharp bend observed by Lai
{\it et al.}\ at $\sigma_0(n)$ persists at finite $T$, fits by the 
piecewise linear function
$f_{\rm plf}(n) = a + b n + c (n - n_{\rm k}) \theta (n - n_{\rm k})$
with the adjustable parameters $a$, $b$, $c$, and $n_{\rm k}$ were 
tried. Here, $n_{\rm k}$ denotes the knee position, and $\theta$ stands 
for the Heaviside step function. For comparison, the data sets were 
approximated by polynomials of third order, which have the same number 
of adjustable parameters. Additionally to the data shown in Fig.\ 1, 
$\sigma(0.47 {\rm K},n)$ and $\sigma(0.87 {\rm K},n)$ were included in
this analysis.

Table 1, left part, presents the optimum knee positions $n_{\rm k}$ and 
the related values of the least square sum $\chi^2$, where weight 1 was 
ascribed to all points. It is remarkable that the same $n_{\rm k}$ 
value, $0.318 \cdot 10^{11}\ {\rm cm}^{-2}$, results for all 
temperatures but 0.47 and $0.35\ {\rm K}$. This $n_{\rm k}$ value is 
only very slightly lower than the abscissa of the knee for the 
$T \rightarrow 0$ extrapolation $\sigma_0(n)$ in Fig.\ 2 of Ref.\ 
\cite{Lai.etal.07}. Moreover, note that, for the four highest $T$ 
values, piecewise linear functions approximate the experimental data 
clearly better than the polynomials.

The advantage of piecewise linear functions over polynomials of third 
order diminishes with decreasing $T$. This trend is not surprising 
because, in the region of activated conduction, 
$\sigma(T = {\rm const.},n)$ vanishes in some exponential way with 
decreasing $n$. Hence, when considering a fixed $n$ range and lowering 
$T$, the $\sigma$ region broadens, and the basic nonlinearity gets 
increasing weight compared to the ``indentation'' at 
$n \approx 0.32 \cdot 10^{11}\ {\rm cm}^{-2}$. Thus, a more meaningful 
comparison of both approximations is obtained restricting all fits to 
comparable $\sigma$ intervals, here to
$\{0.5\,\sigma(T,n_{\rm c}),2.0\,\sigma(T,n_{\rm c})\}$.  Under this
condition, for each of the $T$ values considered, piecewise linear 
functions approximate the data clearly better than polynomials of third 
order, see Tab.\ 1, right part. Moreover, now all fits yield the same 
$n_{\rm k}$ value, $0.318 \cdot 10^{11}\ {\rm cm}^{-2}$.

Thus, for all $T$ values, some characteristic change occurs within a
very small $n$ region close to the knee of the piecewise linear fit. In
this sense, the sharp bend observed by Lai {\it et al.}\ at the
$T \rightarrow 0$ extrapolation $\sigma_0(n)$ turns out to be not
smoothed for finite $T$. In particular, it seems to exist even for
$T \ll T_{\rm F}$, where $\sigma(T,n = {\rm const.})$ is nonlinear and
considerably deviates from the linear extrapolation estimate from the 
vicinity of $T_{\rm F}$. Because 
$n_{\rm k} \approx n_{\rm c}$ for all $T$, it is likely that the bends 
originate from sharp continuous phase transitions separating the regions 
of apparent metallic and activated conduction at finite $T$. 

However, there is a tiny systematic difference between the values of 
$n_{\rm k}$ and $n_{\rm c}$. It presumably arises from the approximation 
by piecewise linear functions being an oversimplification. This aspect 
is considered in the next section.

\begin{center}
\begin{table}
\caption{Results of phenomenological fits of
$\sigma(T = {\rm const.},n)$ data from the Si quantum well study Ref.\
\cite{Lai.etal.07}. Piecewise linear functions (plf) and polynomials of 
third order (pto) are compared. Values of knee positions $n_{\rm k}$
and least square sums $\chi^2$ are given in units of
$10^{11}\ {\rm cm}^{-2}$ and $e^4/h^2$, respectively. The left group of
columns results from consideration of the complete data sets, the right
group is obtained from the data points fulfilling
$0.5\,\sigma(T,n_{\rm c}) < \sigma(T,n) < 2.0\,\sigma(T,n_{\rm c})$.
$N$ denotes the number of data points taken into account.}
\centering
\begin{tabular}{c|cccc|cccc}
\hline
 & \multicolumn{4}{c|}{complete data set}
 & \multicolumn{4}{c}{restricted data set}\\
$T({\rm K})$ & $N$ & $n_{\rm k}$ & $\chi^2_{\rm plf}$ &
$\chi^2_{\rm pto}$ & $N$ & $n_{\rm k}$ & $\chi^2_{\rm plf}$ & 
$\chi^2_{\rm pto}$\\
\hline
4.0 & 15 & 0.318 & 0.010 & 0.024 & 15 & 0.318 & 0.010 & 0.024 \\
2.18 & 15 & 0.318 & 0.006 & 0.019 & 15 & 0.318 & 0.006 & 0.019 \\
1.18 & 15 & 0.318 & 0.010 & 0.019 & 14 & 0.318 & 0.009 & 0.018 \\
0.87 & 15 & 0.318 & 0.016 & 0.023 & 13 & 0.318 & 0.009 & 0.020 \\
0.64 & 15 & 0.318 & 0.028 & 0.031 & 11 & 0.318 & 0.013 & 0.026 \\
0.47 & 15 & 0.316 & 0.061 & 0.050 & 9 & 0.318 & 0.023 & 0.032 \\
0.35 & 15 & 0.315 & 0.097 & 0.063 & 9 & 0.318 & 0.024 & 0.036 \\
\hline
\end{tabular}
\end{table}
\end{center}

\vspace{-0.5cm}

\section{Scaling analysis}

Provided the sharp bend is indeed related to a phase transition, how may 
it arise? What fine structure could it have? On a macroscopic level, 
$\sigma(T,n)$ can be described in the following way. Suppose, on the 
insulating side of the apparent MIT, the $T$ dependence of $\sigma$ 
scales as observed by Kravchenko {\it et al.}\ \cite{Krav.etal.95} at 
the resistivity of MOSFETs, 
\begin{equation}
\sigma(T,n) = \sigma(t)\ \ \ {\rm with}\ \ \ t = T / T_0(n)\ .
\end{equation}
For further references on scaling of $T$ dependences close to the MIT in
various 2d and 3d systems, see \cite{Moe.08}. Assume that this scaling 
holds up to $n_{\rm c}$ where $T_0(n) \rightarrow 0$. Suppose furthermore 
\begin{equation}
T_0(n) = A |\delta n|^{\beta}\ \ \ {\rm with}\ \ \   
\delta n = n - n_{\rm c}\ ,
\end{equation}  
where $A$ is a constant, which might, 
however, be sample dependent. Thus, $\sigma$ should depend only on the
quotient $T/| \delta n |^{\beta}$ or rather on $\delta n/T^{1/\beta}$,
compare Fig.\ 10 of Ref.\ \cite{Krav.etal.95}.

The value of the critical exponent $\beta$ should be universal. It was 
determined by Kravchenko {\it et al.}\ \cite{Krav.etal.95} from their 
MOSFET experiment, $\beta = 1.6 \pm 0.1$. The value 
$n_{\rm c} = 0.322 \cdot 10^{11}\ {\rm cm}^{-2}$ is known from the sign 
change of $\mbox{d}\;\! \sigma / \mbox{d}\;\! T$ as $T \rightarrow 0$. Thus, 
without adjusting any parameter, a scaling check of $\sigma(T,n)$ can be 
performed plotting $\sigma$ as function of $\delta n/T^{1/\beta}$ for 
different $T$. This is done in Fig.\ 2 considering the 
$\sigma(T = {\rm const},n)$ data sets  for $T = 0.35$, 0.47, 0.64, and 
$0.87\ {\rm K}$. Data sets for higher $T$ are not taken into account 
since $\sigma(T,n_{\rm c})$ becomes $T$ dependent above roughly 
$1\ {\rm K}$, presumably because some additional mechanism becomes 
relevant there, see Fig.\ 1 of Ref.\ \cite{Lai.etal.07}.

\begin{figure}
\begin{center}
\includegraphics[width=0.70\linewidth]{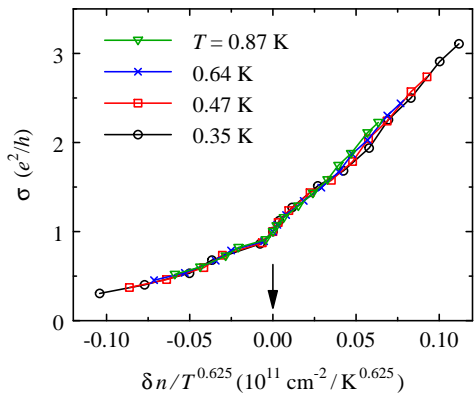}
\end{center}
\caption{Scaling check without adjustment of parameters considering 
the four $\sigma(T = {\rm const},n)$ data sets with $T < 1 {\rm K}$ 
which are analysed in Tab.\ 1.} 
\end{figure}

Note that such a scaling check ascribes higher weight to the points more
distant from $n_{\rm c}$, while the analysis in the previous section is 
particular sensitive with respect to the data very close to $n_{\rm c}$.
Therefore, here, the influence of sample inhomogeneities is reduced in 
comparison to the fits in Tab.\ 1.

Figure 2 shows that, for $n < n_{\rm c}$, all four 
$\sigma(\delta n/T^{0.625})$ curves nicely fall together although no 
parameter has been adjusted. It is unlikely that the agreement of the 
curves arises only by chance, so that Fig.\ 2 can be considered as 
support of the scaling hypothesis. Hence, since $T_0(n) \rightarrow 0$ 
as $n \rightarrow n_{\rm c}$, this graph indirectly indicates phase 
transitions at finite $T$.

Also for $n > n_{\rm c}$, that means within the apparent metallic 
phase, the four curves in Fig.\ 2 collapse. This is surprising: 
In case of conventional metallic conduction, as $T$ vanishes, 
$\sigma(T,n)$ would tend to a finite and $n$ dependent value, 
$\sigma(0,n)$, which increases monotonously with $n$. Vanishing $T$ 
corresponds to diverging $\delta n/T^{0.625}$, so that curves, which 
were drawn for varying $T$ and fixed $\delta n$ in Fig.\ 2, would split, 
in contradiction to the observed scaling. Thus, scaling for 
$n > n_{\rm c}$ cannot be understood in terms of conventional metallic 
behaviour. Superconductivity at $T = 0$ might be an alternative, compare 
the discussion in Ref.\ \cite{Krav.etal.95} and the study of ultrathin 
Bi films by Liu {\it et al.} \cite{Liu.etal.91}.

However, for $n > n_{\rm c}$, a lot of publications have reported 
saturation of $\sigma$ as $T \rightarrow 0$. Thus several questions 
arise: Might the scaling observed here be restricted to the considered 
material or to a tight vicinity of $n_{\rm c}$? What about the influence
of inhomogeneities, sample size or thermal decoupling?

To get a preliminary answer to the first question, reconsider now an 
experimental study of an AlAs quantum well by Papadakis and Shayegan 
\cite{Papa.Shay}. Figure 3 shows scaling checks for data obtained by 
digitising Fig.\ 2 of Ref.\ \cite{Papa.Shay}. To correct for the $T$
dependence of $\sigma$ at $n_{\rm c}$, the quotient
$q(T,n) = \sigma(T,n)/\sigma(T,n_{\rm c})$ is considered here instead of
$\sigma(T,n)$ itself. This approach turned out to be effective for the 
Si quantum well data by Lai {\it et al}.\ in Ref.\ \cite{Moe.08}.

\begin{figure}
\begin{center}
\includegraphics[width=0.70\linewidth]{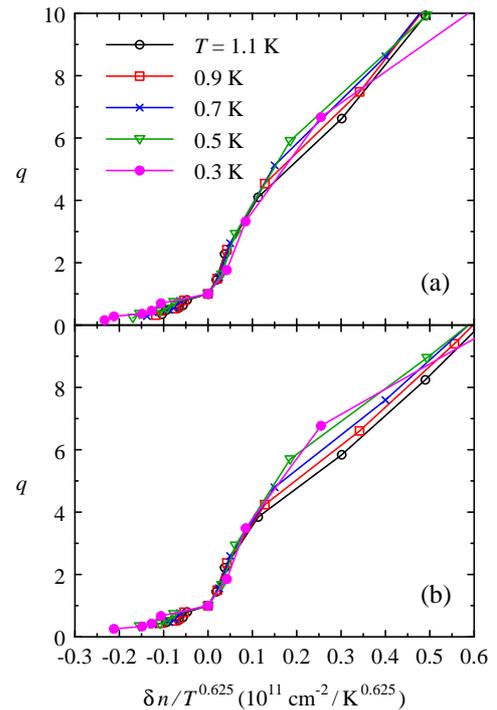}
\end{center}
\caption{Scaling check for AlAs quantum well data from Fig.\ 2 of Ref.\
\cite{Papa.Shay}: (a) high-mobility and (b) low-mobility directions. 
Here, $n_{\rm c} = 0.70 \cdot 10^{11}\ {\rm cm}^{-2}$ is presumed, see
text.}
\end{figure}

The scaling check given in Fig.\ 3 has a hidden parameter: Contrary to
the situation in Fig.\ 2, the $n_{\rm c}$ value has some uncertainty.
According to Fig.\ 2 of Ref.\ \cite{Papa.Shay}, the $n_{\rm c}$ 
values defined by ${\mbox d} \sigma / {\mbox d} T = 0$ as 
$T \rightarrow 0$ are roughly 0.65 and 
$0.70 \cdot 10^{11}\ {\rm cm}^{-2}$ for the high- and low-mobility 
directions, respectively, but $n_{\rm c}$ should be uniquely defined. 
(A possible origin of the direction dependence of the mobility is 
discussed in detail in Ref.\ \cite{Papa.Shay}.) In comparing plots for 
several values of $n_{\rm c}$, the best data collapse was observed for 
$n_{\rm c} = 0.70 \cdot 10^{11}\ {\rm cm}^{-2}$.

Figure 3 (broader $n$ range) resembles Fig.\ 2 to a large extent. Thus, 
although the random dispersion of the data points is here clearly higher
than in Fig.\ 2, the proposed scaling approach might be applicable to 
the AlAs data from Ref.\ \cite{Papa.Shay}. Moreover, also in Fig.\ 3, 
the slope changes significantly at $n = n_{\rm c}$.

Turn now again to the analysis of the Si quantum well data by Lai 
{\it et al}. The ``indentation'' around $\delta n = 0$ in Fig.\ 2 can be
understood in the following way. Since $T_0(n) \rightarrow 0$ and
$t \rightarrow \infty$ as $n  \rightarrow n_{\rm c}$ for arbitrary 
constant $T$, the ansatz 
\begin{equation}
\sigma(t) = \sigma_{\rm c} \cdot (1 - B\, t^{-\nu})
\end{equation}
is suggested. Here $B$ is a dimensionless constant and $\nu$ a positive 
exponent. Hence, 
\begin{equation}
\sigma(T,n) = \sigma_{\rm c} \cdot (1 - C\, |\delta n|^{\beta \nu})
\ \ \ {\rm with}\ \ \  C = B (A/T)^{\nu}\ .  
\end{equation}
Only a rough guess of the value of $\nu$ can be given at present. In 
case of hopping in the Coulomb gap for small $t$, $\nu$ would be 1/2. 
However, according to Ref.\ \cite{Krav.etal.95}, $\nu$ may be expected 
to be smaller for large $t$. In both cases, the product $\beta \nu$ 
would be clearly smaller than 1 so that $\sigma(T = {\rm const.},n)$ 
should have a root-like peculiarity at $n_{\rm c}$. 

This hypothetical peculiarity implies a divergence of 
$\mbox{d}\;\!\sigma / \mbox{d}\;\!n$, and thus also of
$\mbox{d}\;\!\mbox{log}_{10}\;\!\sigma / \mbox{d}\;\!n$. Already at finite
temperature, it indicates the transition between activated and apparent 
metallic conduction. As a consequence, it permits the precise 
identification of the MIT. This approach resembles the strategy used in 
studying ultrathin metal films in Ref.\ \cite{Mack.etal.98}, where the 
crossover between logarithmic and exponential temperature dependence was
related to a change in the thickness dependence of the thickness 
derivative of the conductance.

In reality, the expected divergence would be smoothed to a sharp 
maximum. Fig.\ 3 shows that, indeed,
$\mbox{d}\;\!\mbox{log}_{10}\;\!\sigma / \mbox{d}\;\! n$ exhibits such a 
sharp peak independent of $T$ -- note that a logarithmic scale is used 
to display the derivative values. It is striking, although a natural
consequence of Eq.\ (4), that the peak is always located just at 
$n_{\rm c}$, another indication of a phase transition occurring at 
finite $T$. Of course, the question arises whether or not this sharp 
maximum is specific to the Si quantum well studied in Ref.\ 
\cite{Lai.etal.07}. The comparison of Fig.\ 2 with Fig.\ 3 suggests
that also the AlAs quantum well studied in Ref.\ \cite{Papa.Shay} may 
exhibit such a feature.

\begin{figure}
\begin{center}
\includegraphics[width=0.70\linewidth]{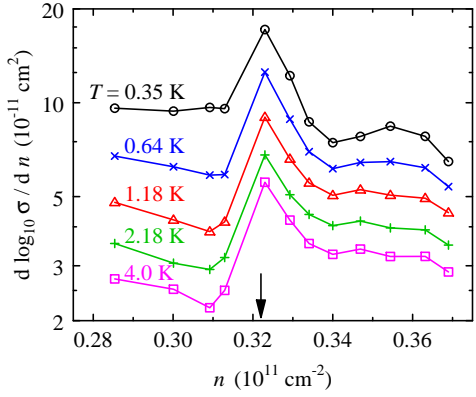}
\end{center}
\caption{$\mbox{d}\;\!\mbox{log}_{10}\,\!\sigma / \mbox{d}\;\!n$ 
versus $n$ for the Si quantum well curves displayed in Fig.\ 1. 
Concerning details of the calculation of the derivative values see Ref.\
\cite{Moe.08}.}
\end{figure}

\section{Challenges}

As a whole, the presented indications give strong support to the 
hypothesis that the MIT at $T = 0$ is connected with a line of sharp
phase transitions at finite $T$. However, this conclusion has to be
confirmed by further experiments. Such studies should primarily focus on
enhancing precision, in particular concerning homogeneity of the 
samples, rather than on reducing the lowest accessible $T$ for the 
following reasons: In plots analogous to Fig.\ 2, the effect of 
inhomogeneities increases with decreasing $T$. For arbitrary $T$, 
inhomogeneities smooth the indentations shown in Figs.\ 1 to 3 and the
peaks in Fig.\ 4.

In case, the hypothetical line of phase transitions at finite $T$ 
exists indeed, three questions arise: Does it terminate at a certain
$T$ value? Under which conditions does scaling hold in the region of
apparent metallic conduction, is there a specific $(T,n)$ region?
What is the nature of the related unconventional phase?

\vspace{0.3cm}

\noindent I am indebted to T. Vojta for an important literature hint.


\end{document}